# Ground based Gamma Ray Studies based on Atmospheric Cherenkov technique at high mountain altitude

A. L. Mishev*, S. Cht Mavrodiev and J. N. Stamenov

*Institute for Nuclear research and Nuclear Energy-Bulgarian Academy of Sciences, 72 Tsarigradsko chaussee, 1784 Sofia, Bulgaria*

*Abstract: In this paper we present a new method for ground based gamma ray astronomy based only on atmospheric Cherenkov light flux analysis. The Cherenkov light flux densities in extensive air showers (EAS) initiated by different primaries are simulated, precisely primary protons in the energy range 100 GeV – 100 PeV, primary gamma quanta in the energy range 10 GeV – 10 PeV and primary iron nuclei in the energy range 10 TeV –100 PeV using the CORSIKA 6 code at high mountain observation level of 536 g/cm$^2$.*

*An approximation of lateral distribution of Cherenkov light in extensive air showers is obtained. The obtained approximation is a nonlinear fit such as Breit-Wigner with few parameters. A detailed study of the energy dependence of the proposed model function parameters is carried out and the fit of model parameters as a function of the primary energy is obtained as well. On the basis of the difference between the model parameters, precisely their behavior as a function of the energy, the strong no linearity of the model, we propose a method, which permits to make the distinction between gamma ray primaries from hadronic primaries. The possible backgrounds for ground based gamma ray astronomy are studied and the efficiency of the method is calculated.*

*Different detector displacements are analyzed using the simulation of simplified mass spectrum of cosmic ray. The detector response is simulated taking into account the physical fluctuation of the processes, the statistical and possible systematic errors. The simulated and reconstructed events are compared and the accuracy in energy and primary mass reconstruction is obtained. Moreover the accuracy in shower axis determination is studied and criteria in shower axis position estimation are proposed.*

## 1.Introduction

The discoveries from space-borne and ground-based instruments have revolutionized the field of gamma-ray astronomy. One of the most exciting challenges in this field is the exploitation in a region of the gap between ground based and space born experiments. Currently gamma-ray energies between 20 and 250 GeV are not accessible to both space-borne detectors,and ground-based air Cherenkov detectors. The scientific potential of the ground based gamma ray astronomy is enormous and covers both astrophysics and fundamental physics. In one hand it is possible to study objects such as supernova remnants, active galactic nuclei and pulsars. On the other the observations especially in the range of low energies will help to understand well the various acceleration mechanisms assumed to be at the origin of very high energy gamma quanta. Several in development projects are based on image technique such as MAGIC [1], HESS [2] and VERITAS [3] are designed to study such type of problems as well. Our aim is to propose an alternative method, also based on atmospheric Cherenkov light registration.The reconstruction of the lateral distribution of Cherenkov light flux in EAS and therefore selection of gamma quanta induced signals from nuclei induced background is the basis of the alternative methods for primary gamma ray study. In the mentioned above energy range taking into account the relatively low densities it is very important to estimate with enough precision the shower axis and thus precisely the energy of the initial primary. Moreover this permits on the basis of complicated analysis to estimate the shape of the lateral distribution of atmospheric Cherenkov light flux densities and therefore to estimate the nature of the initial primary and reject the hadronic events from gamma quanta induced events.

## 2. The Method and Results

The fact that cascades initiated by primary nuclei are different in composition, longitudinal and transverse extension compared to those initiated by primary gamma quanta in the deep atmosphere This results on the corresponding lateral distribution of Cherenkov light flux densities in EAS. At high-mountain altitude the shape of the lateral distributions of Cherenkov light flux densities in EAS initiated by primary gamma quanta and nuclei is more or less similar [4] Fig. 1. Moreover in the energy range of the gap the expected densities of Cherenkov light photons are very low which requires large detectors. At the same time the approximation with nonlinear fit permits even in a case of similar distributions to make the distinction between them. One possibility is using the REGN [5] code, which permits to distinguish even similar distributions as shape and values. For this study one use simulated data. The simulated with CORSIKA 6.003 [6] code using QGSJET [7] and GHEISHA [8] hadronic interaction models actually the lateral distributions of atmospheric Cherenkov light flux densities are seen in Fig.1. The simulations are carried out for high-mountain altitude of 536 g/cm$^2$ in the energy range 10 GeV-10 TeV for primary gamma quanta and 100 GeV – 10 TeV for primary protons.

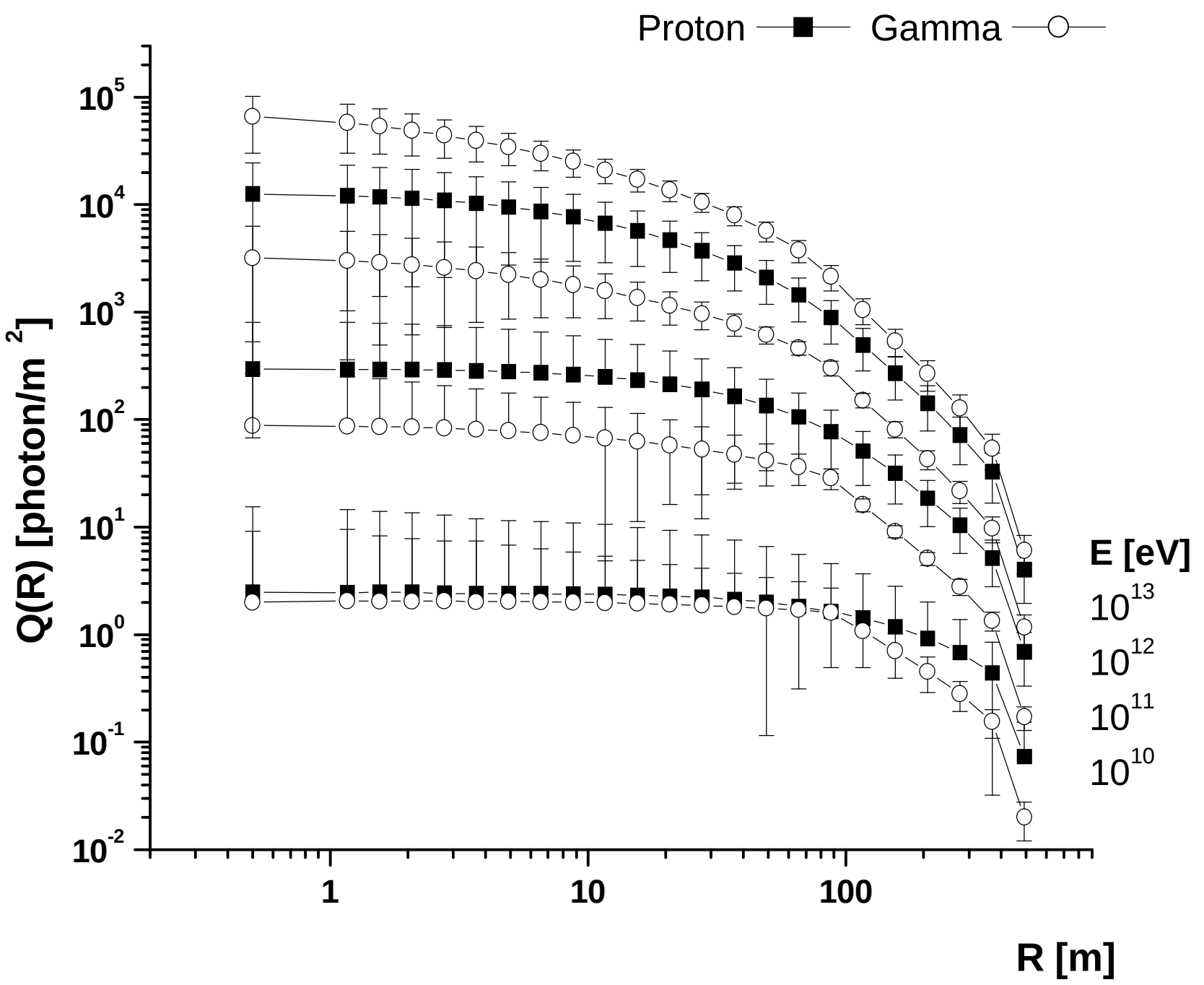

Fig.1 Lateral distribution of atmospheric Cherenkov light flux densities simulated with CORSIKA 6 code (scatter line) and the corresponding approximation (solid line)

* Corresponding author

The differences between lateral distribution of atmospheric Cherenkov light flux densities simulated with CORSIKA 6 code are better seen in fig.2.

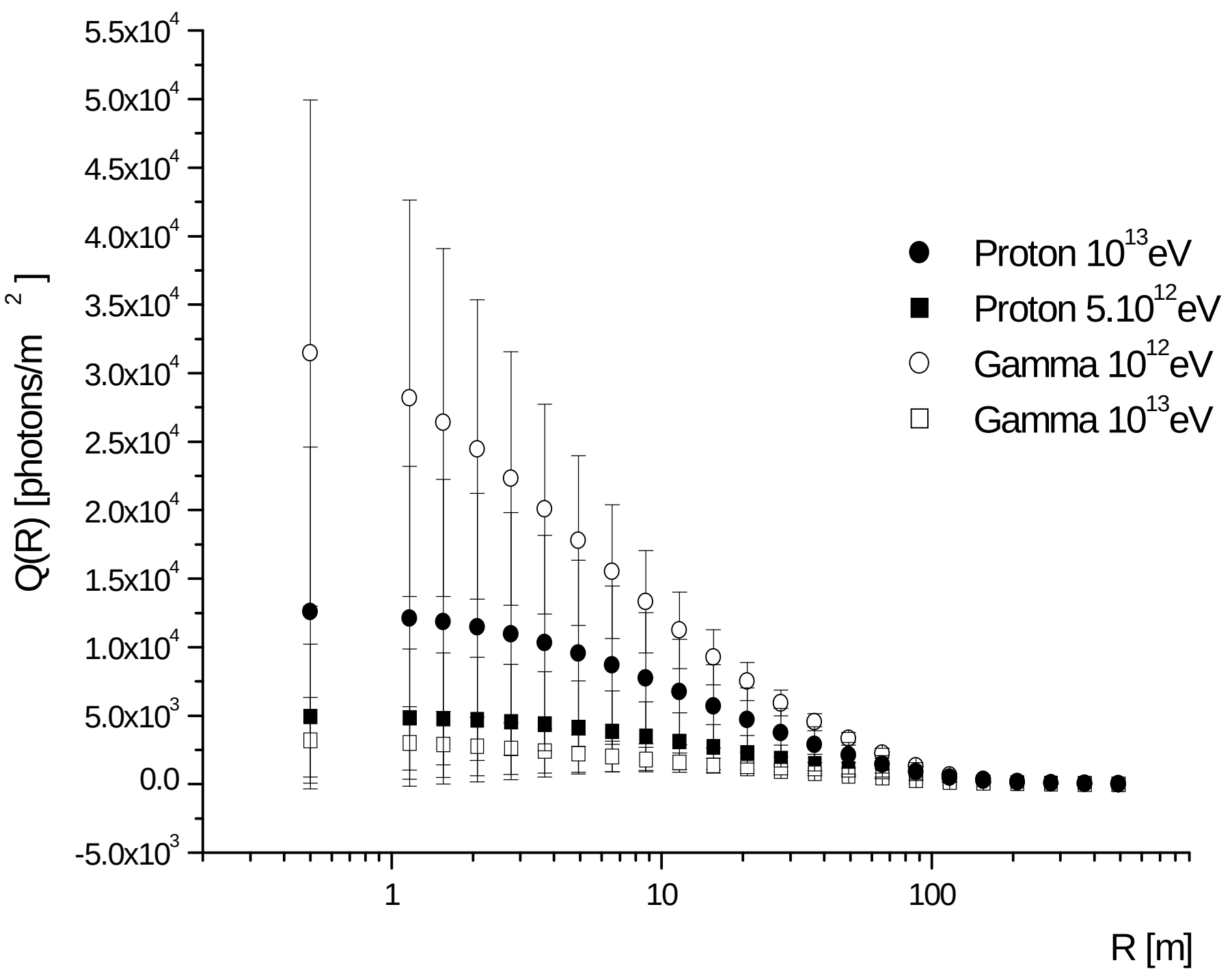

Fig.2 Differences between lateral distribution of atmospheric Cherenkov light flux densities simulated with CORSIKA 6 code for primary gamma-quanta and protons

The obtained lateral distributions are approximated with function such as Breit-Wigner using previously proposed method [9, 10, 11, 12]. The function is the same as the obtained in [13]. The approximation is presented in fig. 1 (solid line). The strong nonlinearity of the model [14, 15, 16] and the obtained monotonic behavior of the model parameters as function of the energy of the initiated primary particle permits to obtain big differences in the $\chi^2$.of the approximation using the same model for the different lateral distribution. The differences in model parameter behavior as a function of the energy and values are shown in fig. 3. This difference is the basis to make the distinction between the initial primaries. In must be pointed out that the solution of the inverse problem based on Gauss-Newton method is iterative and one use the connection between the energy of the initial primary and the total number of Cherenkov photons at given observation level (actually the integral of the obtained lateral distribution function of atmospheric Cherenkov light flux densities). In one hand this relation permits to estimate the energy of the initial primary and on the other hand to adjust the model parameter values as a function of the energy and therefore to distinguish the hadronic initiating particles and gamma-quanta.

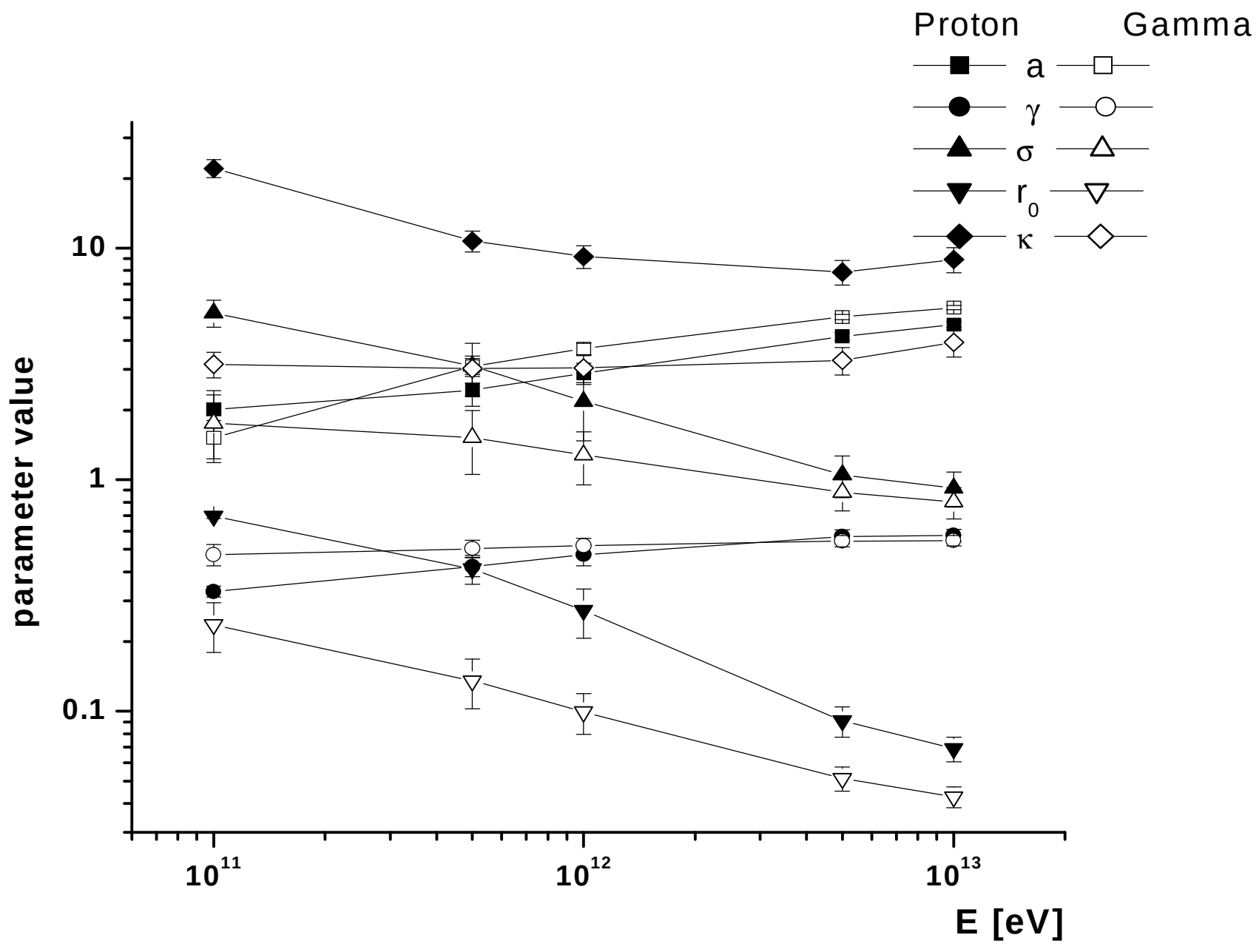

Fig. 3 Difference between model parameter as function of the energy

The obtained approximation gives the possibility for simulation of the detector response of atmospheric Cherenkov telescope.

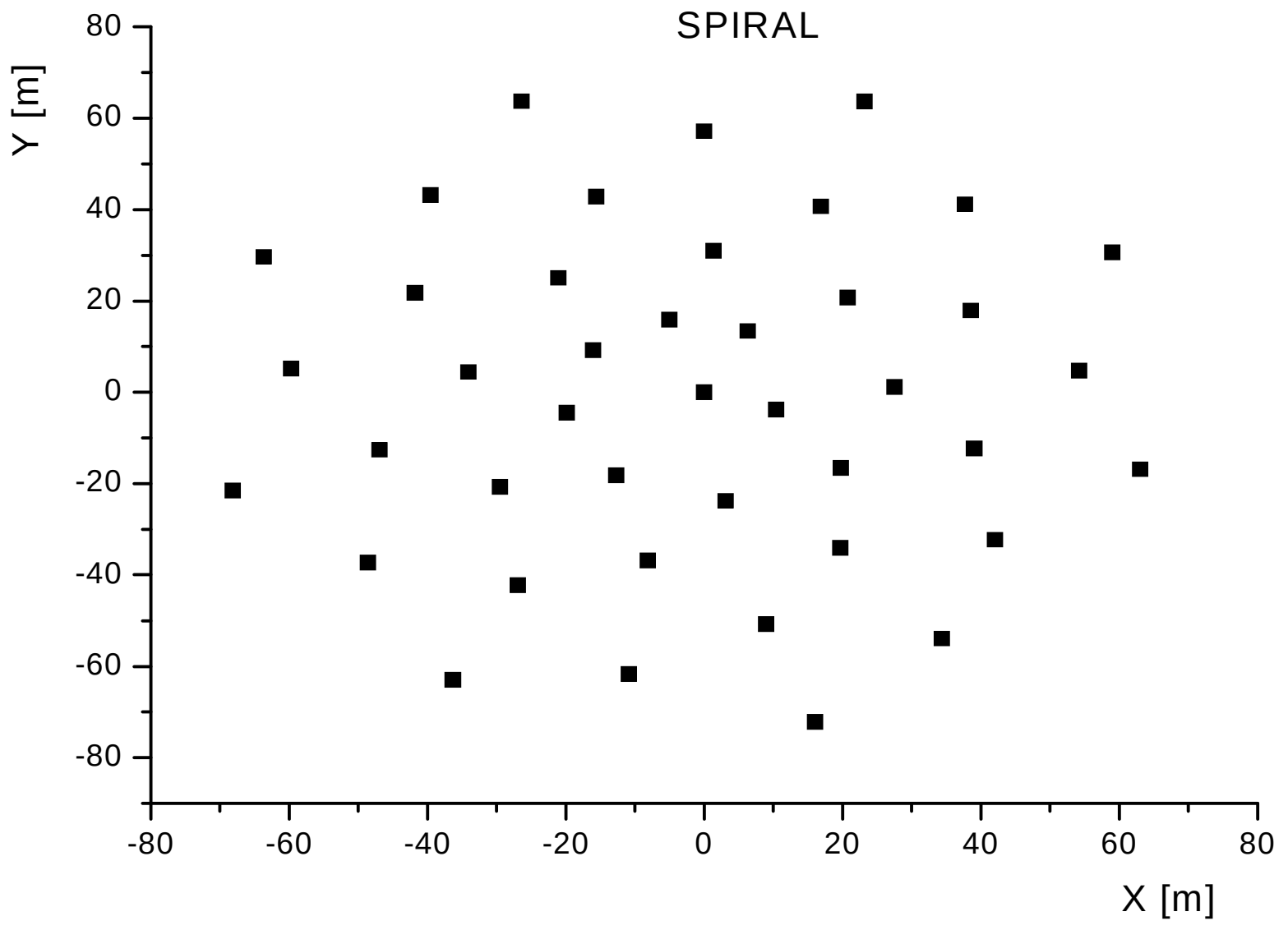

Fig. 4 Atmospheric Cherenkov light telescope with spiral detector displacement

The simulations are carried out for uniform and spiral detector displacement fig. 4. The acceptance of the detector is not taken into account nor the efficiency. The quantity of the Cherenkov photons into the detector is calculated using the obtained approximation and taking into account the detector surface (1 $m^2$). The number of Cherenkov photons in the detector is recalculated, according to the Poisson or Gauss distribution depending of the actual number of Cherenkov photons in the detector. We simulate $10^5$ events (protons and gamma-quanta) according the steep energy spectra and uniform shower axis distribution in the detector field. Using the REGN code and solving inverse problem with input data the simulated showers we estimate the energy and the type of the initial primary. The average of reconstructed events is near to 80 percent. The obtained energy estimation accuracy is 25 percent. This result is comparable with previous obtained results [17] (see fig. 5).

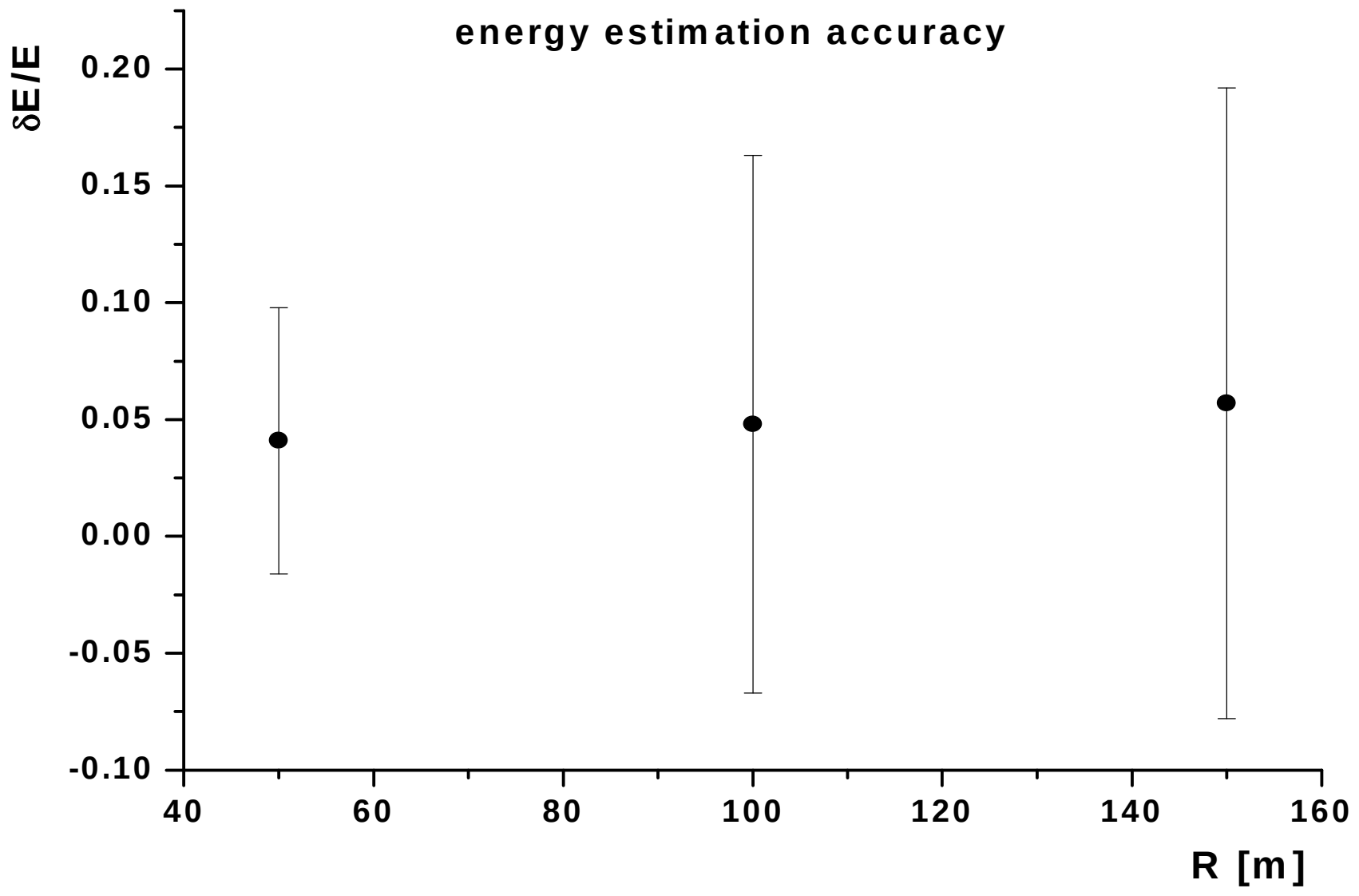

Fig. 5 Energy estimation accuracy for reconstructed events (protons and gamma-quanta) for spiral detector configuration

Similar results are obtained for the uniform detector set but with more detectors. In any case the average of the obtained accuracy and efficiency of the method (average of reconstructed events is quite similar).

**3. Conclusions**

The lateral distribution of atmospheric Cherenkov light flux densities was obtained for primary gamma quanta and protons as initial primary using CORSIKA 6.003 code and were approximated with a nonlinear function such as Breit-Wigner. The analytical form and parameter values of the fit were obtained by solving overdetermined system of equations using the REGN code. The obtained solutions permit an estimate of the energy and the nature of the initiating primary particle thus extracting gamma quanta induced events from hadronic ones. This study is only on methodological level and further development of the method for higher energies is expected.

**Acknowledgements**

We warmly acknowledge our colleagues for discussions during this study especially Dr. L. Alexandrov the main author of REGN code and prof. D. Heck the main author of CORSIKA code. We are grate full to the IT division of INRNE for the assistance.